\documentclass[conference]{IEEEtran}
\IEEEoverridecommandlockouts
\usepackage{cite}
\usepackage{amsmath,amssymb,amsfonts}
\usepackage{algorithmic}
\usepackage{graphicx}
\usepackage{textcomp}
\usepackage{xcolor}
\usepackage{array}
\usepackage{booktabs}
\usepackage{multirow}

\def\BibTeX{{\rm B\kern-.05em{\sc i\kern-.025em b}\kern-.08em
    T\kern-.1667em\lower.7ex\hbox{E}\kern-.125emX}}
\begin{document}

\title{Score-informed Music Source Separation: Improving Synthetic-to-real Generalization in Classical Music\\
\thanks{This work was supported by “REPERTORIUM” Project under Grant Agreement 101095065. Horizon Europe. Cluster II. Culture, Creativity and Inclusive Society. Call HORIZON-CL2-2022-HERITAGE-01-02. \\The authors wish to acknowledge CSC—IT Center for Science, Finland, for computational resources.}
}

\author{\IEEEauthorblockN{Eetu Tunturi}
\IEEEauthorblockA{\textit{Audio Research Group} \\
\textit{Tampere University}\\
Tampere, Finland \\
eetu.tunturi@tuni.fi}
\and
\IEEEauthorblockN{David Diaz-Guerra}
\IEEEauthorblockA{\textit{Audio Research Group} \\
\textit{Tampere University}\\
Tampere, Finland \\
david.diaz-guerra@tuni.fi}
\and
\IEEEauthorblockN{Archontis Politis}
\IEEEauthorblockA{\textit{Audio Research Group} \\
\textit{Tampere University}\\
Tampere, Finland \\
archontis.politis@tuni.fi}
\and
\IEEEauthorblockN{Tuomas Virtanen}
\IEEEauthorblockA{\textit{Audio Research Group} \\
\textit{Tampere University}\\
Tampere, Finland \\
tuomas.virtanen@tuni.fi}
}

\maketitle

\begin{abstract}
Music source separation is the task of separating a mixture of instruments into constituent tracks. Music source separation models are typically trained using only audio data, although additional information can be used to improve the model's separation capability. In this paper, we propose two ways of using musical scores to aid music source separation: a score-informed model where the score is concatenated with the magnitude spectrogram of the audio mixture as the input of the model, and a model where we use only the score to calculate the separation mask. We train our models on synthetic data in the SynthSOD dataset and evaluate our methods on the URMP and Aalto anechoic orchestra datasets, comprised of real recordings. The score-informed model improves separation results compared to a baseline approach, but struggles to generalize from synthetic to real data, whereas the score-only model shows a clear improvement in synthetic-to-real generalization.
\end{abstract}


\section{Introduction}
The target of music source separation is to isolate the signals of individual sources from a mixture. In monaural source separation, tracks are separated from a single-channel input mixture. Music source separation has applications in music upmixing, remixing, virtual reality, and music analysis.

Most of the time, music source separation models use only the mixture of all the instruments as their input. The mixture can contain multiple sources that share timbral characteristics, leading to high correlations among the sources. Additional information can be used to aid the separation model, such as visual information \cite{b10}, instrument activity labels \cite{b11}, or musical score \cite{b3} \cite{b4}. This study focuses on using scores as additional information. It is reasonable to assume that one could have the score available in a real application, especially in the case of classical music. When aligned with the audio, scores indicate the onsets and the offsets of every instrument and the pitch of every note, which can help us know in which frequencies we can expect to find said notes. Score information has been found to improve separation results in frameworks like non-negative matrix factorization (NMF) \cite{b6}\cite{b7}\cite{b8} and hidden Markov models (HMMs) \cite{b9}. 

In the past decade, deep neural networks have provided a significant improvement over NMF and HMMs in music source separation\cite{b17} \cite{b18} \cite{b19} \cite{b20} \cite{b21}. However, there is not much research about integrating score information into deep learning methods. In \cite{b3}, the scores are used to filter the magnitude spectrograms that are used to train a CNN. A score-filtered spectrogram is created for each instrument, and the CNN takes the spectrograms of all of the instruments as its input. In \cite{b4}, the scores are used as weak labels to train separation models in an unsupervised way. The scores are used to enforce the information of every note to be in different dimensions of the latent space of an autoencoder, so that this structured latent space can be modified to separate every note from the mixture. The approach is evaluated by separating the right and left-hand notes from piano recordings, but it has not been studied how this approach would work for separating multiple different instruments.

Source separation of classical music is an especially difficult problem, largely due to the limited amount of training data. There are many instruments in classical music, and songs are recorded such that all of the instruments are played at the same time in the same room. Due to the recording setup, ground-truth signals corresponding to isolated instruments cannot be obtained from normal recordings. Some datasets have been specifically recorded with music information retrieval in mind, such as \cite{aalto} and \cite{urmp}, which contain isolated ground truth signals. However, the number of real recordings remains so low that it is difficult to train deep neural networks on them. There are large synthetic datasets like \cite{synthsod} and \cite{ensembleset}, but based on their baseline results, models trained on synthetic data do not generalize well to real data.

This paper proposes two different ways of using the score information for music source separation: concatenating it to the audio information and using only the score to calculate the separation masks. In the latter approach, the separation masks are less dependent on training data, which helps it generalize better to unseen conditions. We evaluate our methods on the SynthSOD \cite{synthsod}, Aalto anechoic orchestra \cite{aalto}, and URMP \cite{urmp} datasets. Results indicate that the use of score information improves separation performance and generalization from synthetic to real data.

\section{Methodology}
\subsection{Baseline approach}

\begin{figure}[hb]
\centerline{\includegraphics[scale=0.42]{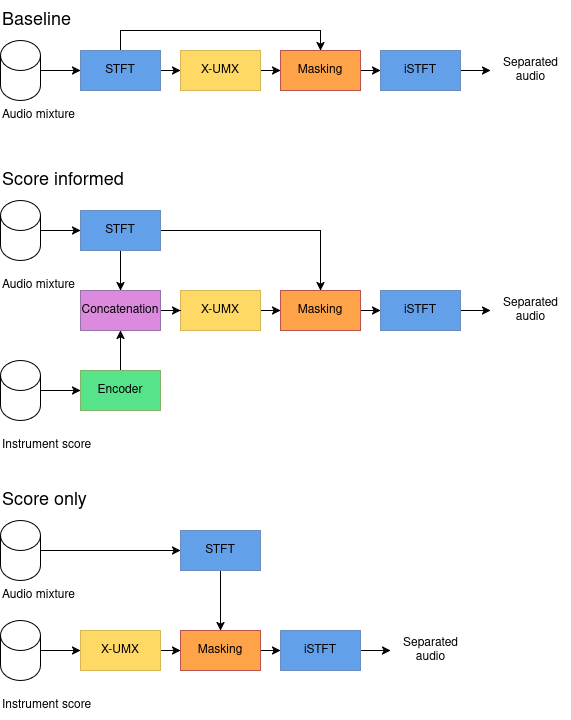}} 
\caption{Block diagrams of the three evaluated methods.}
\label{fig:models}
\end{figure}

Our methods use the magnitude of the STFT of the input mixture to estimate a spectral mask for each target instrument. We apply the spectral masks to the STFT of the mixture to obtain the separated STFTs, and then apply the inverse STFT to obtain the separated signals; this high-level structure of our methods is illustrated in Figure \ref{fig:models}. 

We chose X-UMX \cite{xumx} as our baseline model as it was also used as the baseline for our training dataset \cite{synthsod}. We used the implementation of X-UMX from the open-source Asteroid library \cite{asteroid}. As we trained our models to separate 15 different instruments, the amount of GPU memory required for training became large. Therefore, we split the model into four separate models corresponding to the four instrument families: strings (violin, viola, cello, bass), woodwinds (flute, clarinet, oboe, bassoon), brass (horn, trombone, tuba, trumpet), and percussion (timpani, harp, untuned percussion). The four models are trained completely independently.

X-UMX consists of an encoder, three bidirectional LSTM layers, a decoder, and two global averaging layers. The encoder is a linear layer followed by batch normalization and hyperbolic tangent activation. The decoder contains two linear layers, each followed by batch normalization, and the first followed by ReLU activation. The encoder, LSTM block, and decoder can have multiple branches which all have their own set of weights. We set the number of branches to the number of instruments for each model and for each of the three parts of the model. The decoder has to have as many branches as instruments to obtain the correct number of separation masks, but the encoder and LSTM block could have a different number of branches. However, we decided to keep them all the same for simplicity. There is an averaging layer between the encoder and the LSTM block, and between the LSTM block and the decoder, which computes the mean across the branches. In the baseline model, the input of each encoder branch is the same: the magnitude spectrogram of the audio mixture.
 
For our loss function, we adopt both multi-domain loss and combination loss as described in \cite{xumx}. Multi-domain loss is the sum of the mean squared error between the estimated and ground truth magnitude spectrograms for the frequency domain component, and a weighted signal-to-distortion ratio for the time domain component. In combination loss, instead of only considering the loss of each instrument independently, we also calculate the loss for combinations of instruments by combining their separation masks. The final loss is the mean of all instrument combinations, except for the combination of all instruments together.

\subsection{Proposed approaches}

\begin{figure}[b]
\centerline{\includegraphics[width=90mm]{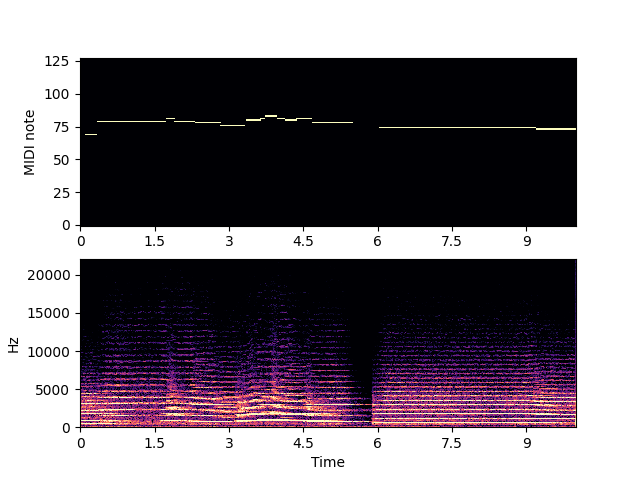}} 
\caption{The piano roll score representation (top) and magnitude spectrogram (bottom) of a 10-second segment of violin from the SynthSOD dataset.}
\label{fig:repr}
\end{figure}

\begin{table*}[ht]
\caption{Signal to distortion ratios [dB] for models trained with SynthSOD and evaluated in the ensembles (up to 5 instruments) of the test partition of SynthSOD, and URMP. The first column of every evaluation dataset indicates the SDR of the original mixtures.}
\label{tab:ensemble}
\scriptsize
\centering
\resizebox{\width}{!}{
\begin{tabular}{rrrrrrrrr}
    \toprule
    Evaluation on:  & \multicolumn{4}{c}{Ensembles in SynthSOD}                             & \multicolumn{4}{c}{URMP}                            \\
    \cmidrule(lr){2-5} \cmidrule(lr){6-9}
     Instrument     & Original   & Baseline  & Score informed  & Score only        & Original    & Baseline     & Score informed  & Score only   \\
    \midrule                                                                                              
     Violin         &  -3.73    &  8.15	    &  \textbf{8.61}  &  5.26             &  -2.32     &  0.70       &  0.72       &  \textbf{5.49}      \\
     Viola          &  -5.87    &  6.06	    &  \textbf{6.58}   &  5.08             &  -5.59     &  0.21       &  0.60       &  \textbf{6.16}      \\
     Cello          &  1.25	   &\textbf{10.28}&  10.06        &  8.63             &  -5.06     &  3.39       &  4.97       &  \textbf{5.83}      \\
     Bass           &  -5.17    &  6.85	    &  \textbf{7.37}   &  6.27             &  -6.56     &  3.43       &  3.36       &  \textbf{6.20}      \\
    \cmidrule(r){1-1} \cmidrule(lr){2-5} \cmidrule(lr){6-9}    
     Flute          &  -11.79   &  2.14      &  \textbf{4.13}  &  1.27             &  -2.67     &  1.19       &  0.95       &  \textbf{3.35}      \\
     Clarinet       &  -7.41	   &  2.36      &  \textbf{4.98}  &  4.32             &  -4.44     &  0.20       &  0.56       &  \textbf{3.78}      \\
     Oboe           &  -5.19	   &  9.81      &  \textbf{10.14} &  3.77             &  -6.54     &  0.17       &  0.31       &  \textbf{1.87}      \\
     Bassoon        &  -3.79	   &  5.77      &  \textbf{7.16}  &  2.07             &  -3.39     &  0.65       &  0.49       &  \textbf{3.04}      \\
    \cmidrule(r){1-1} \cmidrule(lr){2-5} \cmidrule(lr){6-9}    
     Horn           &  -4.38    &  1.51      &  \textbf{4.64}  &  2.10             &  -6.38     &  1.56       &  1.51       &  \textbf{2.84}      \\
     Trumpet        &  -0.02    &  8.27      &  \textbf{9.01}  &  6.25             &  -2.38     &  1.84       &  3.07       &  \textbf{5.73}      \\
     Trombone       &  1.02	   &  7.36      &  \textbf{9.05}  &  6.06             &  -3.84     &  0.69       &  1.83       &  \textbf{3.79}     \\
     Tuba           &  5.08	   &  4.00      &  \textbf{7.85}  &  6.78             &  -6.43     &  0.03       &  1.09       &  \textbf{5.78}     \\
    \cmidrule(r){1-1} \cmidrule(lr){2-5} \cmidrule(lr){6-9}    
     Harp           &  -11.56   &  2.72      &  \textbf{3.40}  &  2.53             &            &         &            &        \\
     Timpani        &  -13.80   &\textbf{3.64}&  0.82          &  0.42             &            &         &            &        \\
     Unt. perc.     &           &            &                 &                   &            &         &            &        \\
    \midrule
     MEAN           & -4.67     & 5.64  & \textbf{6.70}        & 4.34              & -4.63      & 1.17        & 1.62           & \textbf{4.49}       \\
\end{tabular}
}
\end{table*}

We preprocess the score into a piano roll representation. Using the same temporal resolution as the spectrogram, for every time frame we represent the note activity as a vector with length 128 (i.e., the number of MIDI notes), where the note activity is denoted by 1 (on) or 0 (off). Figure \ref{fig:repr} shows the piano roll of one of the instruments of a track from SynthSOD aligned with the corresponding magnitude spectrogram. Before feeding the score to the model, it is aligned with the audio during preprocessing, as explained in Section \ref{align}.

In our score-informed model, we first encode the score information with a small encoder consisting of a 3-layer bidirectional LSTM followed by batch normalization and ReLU activation. The score encoder has its own set of trainable weights for each instrument. The encoded score information is concatenated framewise with the magnitude spectrogram of the mixture and then given to an X-UMX model with the same hyperparameters as the baseline, with the exception of the input size. Each branch of the encoder of the X-UMX gets the score of one instrument. We also tried different approaches to integrate the score information with the audio, but we obtained very similar results with all of these approaches. The different approaches we experimented with included combining the score and audio at different intermediate layers in the X-UMX architecture, and multiplying the score and audio instead of concatenating them.

We also propose a score-only model. The score-only model uses only the score to create the spectral masks, as seen in Figure \ref{fig:models}. In this case, we do not use a separate encoder for the scores, instead, X-UMX directly takes the piano rolls of every instrument as inputs, each in its own encoder branch. The score representation used is the same as for the score-informed model. For the score-only model, we removed the input and output normalizations of X-UMX since we found that it worked better without them in this case.

\subsection{Score alignment} \label{align}
It is important for separation performance that the score and audio are aligned. Of the three datasets used in our experiments, URMP \cite{urmp} and the Aalto anechoic orchestra dataset \cite{aalto} provide aligned scores. However, our training dataset, SynthSOD \cite{synthsod}, does not provide aligned score information. SynthSOD is a large-scale dataset of synthetic classical music containing both ensembles and orchestral music. We obtained our score information from the MIDI files that were used to synthesize the dataset. However, since random tempo segments were used to increase the diversity of the dataset, there are significant misalignments between the original scores and the audio. Misalignments could cause problems in training the model, so we chose to align the MIDI files with the audio as a preprocessing step. 

To align the scores to the audio, we used the system described in \cite{aligment}. This system splits the score into different segments according to the concurrent notes and their transitions, which are synthesized and modeled using non-negative matrix factorization, and aligns them to the audio using a dynamic time warping algorithm.

From the alignment, we obtain a CSV file for each song, which contains, for every note in the song, the corresponding onset, offset, pitch, and instrument. The synthesizer used for the alignment cannot parse some of the MIDI files perfectly, so we chose to exclude those songs from all of our experiments. We release the aligned scores and updated metadata files\footnote{[Online]. Available: https://doi.org/10.5281/zenodo.15575778}. In this paper we focus on the effectiveness of aligned scores. Temporal misalignments will affect results, but we leave the study of those effects for future work.

\begin{table*}[ht]
\caption{Signal to distortion ratios [dB] for models trained with SynthSOD and evaluated in the orchestra pieces (more than 5 instruments) of the test partition of SynthSOD, and Aalto. The first column of every evaluation dataset indicates the SDR of the original mixtures.}
\label{tab:orchestra}
\scriptsize
\centering
\resizebox{\width}{!}{
\begin{tabular}{rrrrrrrrr}
    \toprule
    Evaluation on:  & \multicolumn{4}{c}{Orchestras in SynthSOD}                             & \multicolumn{4}{c}{Aalto}                                            \\
    \cmidrule(lr){2-5} \cmidrule(lr){6-9}
     Instrument     & Original    & Baseline   & Score informed  & Score only         & Original    & Baseline     & Score informed  & Score only    \\
    \midrule                                                                                             
     Violin         &  -9.05	     &\textbf{4.07}&  3.91        &  2.12                &  -7.27      &  2.60	      &\textbf{2.72}    & 2.09      \\
     Viola          &  -10.61     &  1.87      &\textbf{2.60}  &  2.10                &  -13.36     &  0.01	      &\textbf{0.25}    & -0.07      \\
     Cello          &  -5.60	     &  3.86      &\textbf{4.24}  &  3.53                &  -15.07     &  -2.68       &\textbf{-0.42}   & -1.33      \\
     Bass           &  -5.65	     &  7.58      &\textbf{8.04}  &  7.44                &  -15.63     &  4.74	      &  5.04	        & \textbf{6.08}      \\
    \cmidrule(r){1-1} \cmidrule(lr){2-5} \cmidrule(lr){6-9}  
     Flute          &  -12.60     &  1.30      &\textbf{3.50}  &  2.06                &  -15.32     &  0.06        &\textbf{0.45}    & -1.33       \\
     Clarinet       &  -8.95	     &  0.45      &\textbf{2.64}  &  1.69                &  -8.31	   &  0.12        &  0.54           &\textbf{1.62}     \\
     Oboe           &  -10.51     &  4.96      &\textbf{5.52}  &  3.19                &  -5.96	   &  0.96        &  0.92           &\textbf{2.59}     \\
     Bassoon        &  -13.29     &  0.69      &\textbf{1.79}  &  1.36                &  -7.99	   &  1.06        &  0.43           &\textbf{2.19}     \\
    \cmidrule(r){1-1} \cmidrule(lr){2-5} \cmidrule(lr){6-9}
     Horn           &  -5.85	     &  1.23      &\textbf{2.98}  & 1.64                 &  -5.98      &  0.77	      &  0.46	        &\textbf{1.83}       \\
     Trumpet        &  -6.23	     &  2.21      &\textbf{3.19}  & 2.97                 &  -14.49     &\textbf{-0.02}&\textbf{-0.02}   & -3.04         \\
     Trombone       &  -11.89     &\textbf{0.00}&  -0.02       & -0.91                &       &        &           &         \\
     Tuba           &             &            &               &                      &       &        &           &         \\
    \cmidrule(r){1-1} \cmidrule(lr){2-5} \cmidrule(lr){6-9}
     Harp           &  -12.24     &  0.00      &\textbf{0.72}  & 0.18                 &       &        &           &               \\
     Timpani        &  -11.50     &\textbf{3.18}&  1.12	      & 1.15                 &       &        &           &               \\
     Unt. perc.     &  -17.93     &  7.54      &  -2.52        &\textbf{8.64}         &       &        &           &               \\
    \midrule
     MEAN           &  -10.14     &\textbf{2.78}&  2.69        & 2.65                 & -10.94      & 0.76       & 1.04          &\textbf{1.06}       \\
\end{tabular}
}
\end{table*}

\section{Experiments}
\subsection{Training setup}
We train all of our models on the train partition of SynthSOD, excluding the songs that we were unable to align from the training set, as mentioned in Section \ref{align}. Of the 408 original songs, we were able to successfully align 351. SynthSOD contains two different violin tracks for many songs, which we chose to combine into one single violin track. As done in the original baseline of SynthSOD, we also join the piccolo to the flute and the coranglais to the oboe, because the piccolo and coranglais have little data in the dataset and are very similar to the flute and oboe, respectively. All audio in SynthSOD has a sampling rate of 44.1 kHz, which we also use for training. The STFT was computed using a window size of 4096 with a hop size of 1024. We train using 6-second segments of audio and score, which are randomly sampled during the training. 

\subsection{Evaluation setup} \label{evalsetup}
For evaluation, we use three datasets: the test partition of SynthSOD \cite{synthsod}, the Aalto anechoic orchestra dataset \cite{aalto}, and the University of Rochester Multi-Modal Music Performance (URMP) dataset \cite{urmp}. SynthSOD contains a mixture of ensembles, which are simpler songs with fewer instruments, and orchestral pieces, which contain more instruments. To make the results more comparable to the other datasets, we split the SynthSOD test partition into ensembles (5 or fewer instruments) and orchestral pieces (more than 5 instruments) and show the results separately. Even though we show the results separately, all our models have been trained with both ensembles and orchestral pieces. 

The Aalto dataset contains four real recordings of symphonic music, for which note onset and offset annotations are provided in CSV format. The URMP dataset consists of 44 real recordings of songs with between two and five instruments each. We have omitted songs that only contain one kind of instrument from the evaluation. We also omit songs that contain saxophone, since it was not in our training data. URMP audio has a sampling rate of 48 kHz, so we resample it to 44.1 kHz to match the other datasets. URMP provides note onset, frequency, and duration annotations in a CSV format for each individual track. We noticed that every track of the URMP recordings contains a distinct low-frequency noise, which could impact the evaluation results. Therefore, we decided to denoise the recordings using Wiener filtering based on noise statistics calculated from silent regions of the recordings. All of the evaluations in this paper were done using the denoised tracks. We provide the Python script used for the denoising in our Github repository\footnote{[Online]. Available: https://github.com/ee7u/score-mss}, along with the code used to train and evaluate the models.

We use signal-to-distortion ratio (SDR) as a metric to evaluate the separation performance. We use the museval library for computing the SDRs \cite{museval}. The SDRs are computed in non-overlapping one-second frames for each instrument. We take the median of the frame-wise SDRs to obtain song-level metrics, and then we take the median of the songs to get the final results. The museval library excludes silent frames from the metrics. Museval defines silent frames as frames where all of the samples are exactly zero. However, for the URMP and Aalto datasets, the silent parts are not exactly zero. Before the evaluation, we take each of the ground-truth tracks and set all of the silent one-second windows to exact zeros, considering a window as silent if the absolute values of all of the samples in the window are below 0.01. We also applied this threshold to the SynthSOD test partition, which explains why the results of the baseline presented in this paper are better than the ones presented in the original SynthSOD paper. However, these results better represent the separation performance of the models since SDR diverges when the target energy approaches zero. Silent frames were having a disproportionate impact on the final SDR results despite being perceptually negligible.

\section{Results}

We have retrained the baseline model for a fairer comparison since we had to omit some of the data due to the alignment problems described in Section \ref{align}. Table \ref{tab:ensemble} shows the SDRs for the ensembles in the test partition of SynthSOD and the URMP dataset. In SynthSOD, the score-informed model achieves 6.70 SDR, which is 1.06 dB better than the baseline's 5.64 SDR. The score-only model's 4.34 SDR is significantly worse than the baseline, which could be expected since the model has less information at its input. In the URMP results, the score-informed model beats the baseline as well, with the score-informed model achieving 1.62 SDR, which is 0.45 dB better than the baseline's 1.17 SDR, but it presents the same generalization issues as the original audio-only baseline. However, the score-only model does clearly better in URMP, achieving 4.49 SDR, which is a 3.32 dB improvement over the baseline, while having the best metrics in all instruments. The score-only model's performance is almost the same in both the synthetic and the real-recording domains, with only a small difference of 0.15 dB between the two datasets. This shows the score-only model's ability to generalize from synthetic data to real data without overfitting to the specific audio characteristics of the synthetic data since it does not use the audio to compute the frequency masks of every instrument. 

Table \ref{tab:orchestra} contains the results for orchestras in the test partition of the SynthSOD dataset and the Aalto dataset. These datasets are much more challenging, which can be seen in the results. In orchestras in SynthSOD, all of the models have similar performance, as all of them fall within only 0.13 dB of one another. Looking at the instrument specific metrics, the score-informed model seems to be the most consistent, but the one outlier score for untuned percussion brings down the mean score a lot. Similarly, in the Aalto dataset, all of the models are separated only by 0.3 dB. The score-only model is the most consistent, but the results for all instruments except bass are bad. Both of our methods improve slightly compared to the baseline in Aalto, but in all of our experiments, the performance of the models in both matched and mismatched conditions indicates that the separation of orchestral pieces remains a challenging task. 

\section{Conclusion}
We presented two ways of incorporating score information into deep neural networks for music source separation. The score-informed model, which uses the concatenation of the audio mixture and the score to create the separation masks, shows a slight improvement over the baseline in both synthetic and real data. We also show that the score-only model, which uses only score information to create the separation masks, can generalize to real data with minimal overfitting. The score-only model clearly beats the baseline model across the board in all instruments by a mean of 3.32 dB SDR in the URMP dataset. The availability of datasets with real recordings is one of the biggest problems in music source separation, so the generalization capability of the score-only model presents a promising direction for future work.

\end{document}